## Microquasar Jet Irradiation of the Proto-Solar Nebula?

Yousaf M. Butt[1] & Nikos Prantzos[2]

[1] *Harvard-Smithsonian Center for Astrophysics, 60 Garden St., Cambridge, MA 02138, USA.*

[2] *Institut d'Astrophysique de Paris, 98bis Bd Arago, Paris, France.*

### Abstract

We explore the possibility that a now-extinct microquasar may have irradiated the proto-solar neighborhood, causing the 'anomalously' high local $^{11}B/^{10}B$ isotopic ratio. Given the population and typical lifetimes of known radio-emitting X-ray binaries we find the probability of such an event having occured is not unreasonable. We comment on some tests of the scenario that could be carried out by observing the elemental abundances in the vicinity of microquasars, in particular SS433.

### Introduction

Though broadly successful, the Standard Galactic Cosmic Ray (SGCR) picture of light element nucleosynthesis cannot explain the observed abundance ratios of $^7Li/^6Li$ (~12) and $^{11}B/^{10}B$ (~4) measured in meteorites. The theoretically calculated SGCR values, resulting by folding the CNO spallation cross-sections with SGCR particle spectra, are ~1.5 and ~2.5 respectively (e.g. Reeves, 1994). Several additional stellar sources of $^7Li$ have been proposed that can satisfactorily explain the high $^7Li/^6Li$ ratio (eg. via the fusion of $^3$He and $^4$He in AGB stars or novae). In the case of the boron isotopic ratio, the only appealing solution up to now is neutrino-induced spallation of $^{12}$C in type-II supernovae, which may provide additional $^{11}$B (the 'nu-process', Woosley et al., 1990). However, persisting uncertainties in supernova neutrino spectra do not allow any definitive conclusions regarding the validity of this scenario (see, e.g., Yoshida et al. 2004 and references therein).

We propose an alternative solution – *in addition to, or instead of, the nu-process*: the $^{11}$B/$^{10}$B ratio could have been boosted over the SCGR value by an energetic baryonic jet from a local microquasar system impacting the parent molecular cloud core of the proto-solar nebula ~4.6 billion years ago.

A microquasar is a stellar-mass black hole or neutron star that accretes material from a regular companion star and ejects twin opposed jets of material at high (usually relativistic) speeds (eg. Mirabel & Rodriguez, 1999). Recent work indicates that at least some microquasar jets contain nuclei, and not just an electron-positron plasma (eg. Gallo et al., 2005; Heinz 2006; Brinkman, Kotani & Kawai, 2005), and that they may be much more powerful than their radiative signatures let on (Fender, Maccarone and van Kesteren, 2005; Heinz and Grimm, 2005). Such powerful baryonic (nuclei-rich) jets will induce nuclear reactions on the ambient interstellar material (ISM) altering its isotopic make-up. Here we argue that a baryonic microquasar jet irradiating the proto-solar nebula ~4.6 billions years ago can fully explain the meteoritic $^{11}$B/$^{10}$B isotopic ratio, which is significantly lager than predicted by the standard model of light element nucleosynthesis.

## 2. Irradiation of the proto-solar cloud core by a baryonic jet

An inspection of the CNO spallation cross-sections shows that the only energy range where a high $^{11}$B/$^{10}$B ratio can be produced *without an associated overproduction* *of* $^6$Li (from alpha+alpha fusion) is the narrow range slightly below E~10 MeV/nucleon (Figure 1), which also lies just above the threshold of most spallation reactions producing Li, Be and B (eg. Ramaty et al., 1997). For our calculations we will assume monoenergetic jets of initial energy ~10MeV/nucleon. Although this value is towards the low end of the range of measured microquasar jet speeds, it is roughly compatible with that of LS5039 (Paredes et al., 2005). Moreover, considering that the sample of Galactic microquasars whose jet speeds have been measured is far from complete (and may be subject to selection effects favoring the higher jet speeds), assuming such a jet speed is not unreasonable. We adopt a conservative value of P=$10^{37}$ erg/s for the jet power, which is two orders of magnitude less than that of SS433 (Brinkmann et al. 2005); the jet intensity is then J~P/E~$10^{42}$ s$^{-1}$.

Molecular cloud cores (MCC) are the densest regions inside molecular clouds and constitute the birthplaces of stars. They have densities n ~$10^4$-$10^5$ cm$^{-3}$, sizes *l*~0.1-1 pc and masses from a few tens to a few thousand M$_\odot$ (eg. Rathborne et al., 2006). For the purpose of our discussion, we adopt n =$10^4$ cm$^{-3}$, *l*=0.5 pc~1.5 $10^{18}$ cm and M~ n *l*$^3$ ~ 25 M$_\odot$.

Particles with energies E~10 MeV traversing hydrogen suffer ionization losses dE/dx~100 MeV g$^{-1}$ cm$^{-2}$ (eg. Barkas et al., 1964). Assuming the jet beam traverses the MCC in nearly straight line (i.e. assuming negligible diffusion to magnetic field inhomogeneities), it goes through an effective path length $\Delta$x ~ n m$_p$ *l* ~0.023 g cm$^{-2}$ (where m$_p$ is the proton mass) and consecutively its nuclei (protons and alphas) lose an energy $\Delta$E = dE/dx $\Delta$x ~2 MeV/nucleon, i.e. a small fraction of their initial energy. Thus, for all practical purposes, the energy of the beam particles may be considered as approximately constant E ~9 MeV/nucleon during its propagation through the MCC; in those conditions (ie. thin target approximation) no detailed transport for the beam particles inside the MCC needs to be considered.

During the beam propagation, its protons and alpha particles spallate the CNO nuclei of the MCC. Assuming solar composition ($n_{CNO}/n_H \sim 10^{-3}$) for the MCC, the number density of its CNO is $n_{CNO} \sim 10$ cm$^{-3}$. In the thin target approximation the number of $^{11}$B producing reactions per incident particle is $\Delta N \sim \sigma\, n_{CNO}\, l \sim 10^{-6}$ (where $\sigma \sim 100$ mb is a typical cross-section for the production of $^{11}$B at E$\sim$9 MeV/nucleon). The total rate of $^{11}$B producing reactions during the beam propagation is then $dN/dt \sim J\,\Delta N$ $\sim 10^{36}$ s$^{-1}$. Considering a conservative effective[1] irradiation time of just $\Delta t_{eff} \sim 3\ 10^5$ yr, the total number of $^{11}$B nuclei produced is $N \sim dN/dt\ \Delta t_{eff} \sim 10^{49}$. This is approximately equal to the number of $^{11}$B nuclei already extant in the 25 M$_\odot$ ($\sim$3 $10^{58}$ H atoms) of the MCC, which are produced by SGCR. The latter is derived by noting that the solar system number fraction $^{11}$B/H=6 $10^{-10}$ (Lodders 2003), but SGCR may account for only $\sim$60% of that, i.e. for $^{11}$B/H=4 $10^{-10}$.

This rough calculation suggests that, with reasonable values for the MCC parameters and the jet properties, the number of $^{11}$B nuclei produced by the jet is of the same magnitude as the number of already extant $^{11}$B nuclei in the MCC (produced by SGCR). A small variation in the MCC parameters can be compensated by a corresponding variation in the jet parameters: e.g. an increase of the mass of the MCC can be compensated by assuming an equal increase for the jet power or the time of irradiation.

---

[1] In general, the irradiation of the MCC is not continuous, since for most natural geometrical configurations the jet can only impact the cloud periodically (due eg. to rotation or precession of the source). Thus, the adopted irradiation time $\Delta t_{eff}$=3 $10^5$ yr is just the effective time of interaction between the jet and the MCC; the true time of interaction may be much longer, by e.g. a factor of ten or more, which is closer to the true active lifetimes of HMXBs such as SS433. The typical lifetimes of Low Mass X-ray Binaries (LMXBs) are even longer: in the range tens-100's of Myrs (eg. Romani 1992).

The gravitational binding energy of the MCC is $V \sim G\, M^2/R \sim 10^{44}$ erg (where $R=l/2$). The total energy of the jet deposited in the MCC is $E \sim 0.2\, P\, \Delta t_{eff} \sim 2\ 10^{49}$ erg, about five orders of magnitude larger than the MCC binding energy. The jet energy is spent in ionization of the MCC gas, the electrons of which recombine rapidly with nuclei by emitting photons with energies up to $\varepsilon \sim 13.6$ eV ($\sim 10^{-11}$ ergs). The mean free path of those photons inside the MCC is $\lambda \sim (n\, \sigma_T)^{-1} \sim 10^{20}$ cm (where $\sigma_T$ is the Thomson cross-section), i.e. much larger than the size of the cloud ; thus the MCC is transparent to the recombination radiation. Obviously, if the MCC can radiate faster than the jet energy is deposited, it can cool efficiently. The recombination rate of $H^+ + e^- \rightarrow H +$ photon is given by $r = k\, n^2$ (assuming complete ionization); the coefficient $k$ is a decreasing function of temperature, with $k > 10^{-12}$ cm$^3$ s$^{-1}$ for T<1000 K (e.g. Verner and Ferland 1996, see their Fig. 7). Thus, even if the cloud were heated up to 1000 K, it would radiate energy locally at a rate $r\, \varepsilon \sim 10^{-15}$ erg cm$^{-3}$ s$^{-1}$. The total power radiated from its volume $V \sim l^3 \sim 10^{54}$ cm$^3$ would be then $Q \sim r\, \varepsilon\, V \sim 10^{39}$ erg s$^{-1}$, i.e. the MCC would cool at a rate about a hundred times larger than the energy deposition rate from the jet. Of course, the MCC would never reach such a high cooling rate: in reality it would be only partially ionized, and its cooling rate would be equal to the heating rate. In any case, its overall stability would not be threatened by the impact of the jet.

### 3. Likelihood of Microquasar Induced Nucleosynthesis

It is thought that all radio-emitting XRBs (REXRBs) are potentially microquasars and 43 such systems are known to exist in the Galaxy – though this number is a strict lower-limit: the true population of microquasars is likely in the few 100's since many have simply not yet been detected (see eg., Paredes 2005, and references therein; also White and van Paradijs, 1996). Importantly, even extinct microquasars have played a role in Galactic nucleosynthesis: if ~300 microquasars are present in the Galaxy at any given moment and their lifetime is $\tau_\mu \sim 10^{7\text{-}8}$ years each (eg.

Romani 1992), then over the $\tau_G$~10Gyr lifetime of the Galaxy there have existed ~300($\tau_G$ / $\tau_\mu$) or several tens of thousands of such systems.

The Galaxy-wide power residing in all currently active microquasar jets may be as high as ~1/10th the total Galactic supernova power, or about equivalent to the GCR power (eg. Fender, Maccarone and van Kesteren, 2005; Heinz and Grimm, 2005). It is noteworthy that the jets of SS433 alone constitute ~10% of the total GCR power (Brinkmann, Kotani and Kawai, 2005) although all of that power is concentrated in roughly monoenergetic jets of ~33.5MeV/n (0.26c). Since such extreme power is liberated in a small volume (the lobes of SS433 have radius ~60 parsec, tiny relative to the size of Galaxy), the local (to SS433) isotopic processing effects from its jets will be tremendous as compared to the 'background' SGCR effects. Determining the light element abundances in the W50 lobes of SS433 would thus be a straightforward and stringent test of our hypothesis. In particular, from Fig 1 (lower left panel), we can see that the most sensitive probe of our scenario will be the $^6$Li/$^9$Be ratio since it varies widely in the 10-70 MeV/n range. Since both these isotopes are exclusively spallation products, finding supersolar values near a known microquasar would essentially prove our hypothesis.

We will examine the Galaxy-wide implications of microquasar jet induced nucleosynthesis in a forthcoming paper (Butt et al., *in preparation*).

For the present purpose of evaluating the likelihood of the possible irradiation of the proto-solar MCC by a microquasar jet, we need to determine the probability of finding such an MCC within range (say, ~60pc, based on SS433) of an active microquasar's jet during the cloud's collapse phase of ~few Myrs. Since a ~few Myrs is less than the typical lifetime of both HMXBs and LMXBs it corresponds to a single generation of actively accreting microquasars which number in the ~few hundreds at any given time (Paredes 2005, and references therein; also White and van Paradijs, 1996). Taking a radius of 15kpc for the Galaxy, and 100pc for its scale height we find that there ought to be an actively accreting microquasar every ~500pc if they are distributed randomly. Thus the probability of finding a MCC within ~60pc of one of these microquasars will be about 1 in 300, which is small but not unreasonably so. Since ~10 stars form in the Galaxy every year, a star whose pre-stellar MCC has been irradiated by a microquasar ought to be formed every ~30 years. (Whether or not such a star will have a solar-system like $^{11}B/^{10}B$ ratio depends of course on the jet speed, and other parameters of the jet as well as the pre-stellar MCC.)

If the proposed scenario is correct then most stars ought to have $^{11}B/^{10}B$~2.5, and the solar system is somewhat (but not very) exceptional.

## 4. Conclusions

We have found that a microquasar irradiating the proto-solar nebula with jets of ~10 MeV/n with an intensity of ~$10^{37}$ ergs/sec for an effective time of just ~300,000 years can naturally explain why the observed local $^{11}B/^{10}B$ ratio is larger than expected from the SCGR picture. Moreover, the scenario we describe is not peculiar to the solar neighbourhood: energetic microquasars – both existing and extinct – throughout our Galaxy have left, and continue to leave, strong nucleosynthetic imprints in their locales and observational searches of these signatures could prove interesting.

## References:

Barkas, W., Berger, M., *Tables of energy losses and ranges of heavy charged particles*, (SP-3013, Washington DC, NASA) 1964

Brinkmann, W., Kotani, T. and Kawai, N., Astronomy and Astrophysics, v.431, p.575-586 (2005)

Fender, R. P.; Maccarone, T. J.; van Kesteren, Z., MNRAS, Volume 360, Issue 3, pp. 1085-1090 (2005)

Gallo, E., Fender, R., Kaiser, C., et al., Nature 436, 819-821 (2005)

Heinz, S., ApJ 636, 316 (2006)

Heinz, S., and Grimm H-J, ApJ 633, 384 (2005)

Lodders, K., ApJL 591, 1220L (2003)

Mirabel, F. and Rodriguez, L. F., Ann.Rev.Astron.Astrophys. 37, 409-443 (1999)

Paredes, J. M., Invited talk presented at the International Symposium "High-Energy Gamma-Ray Astronomy", 26-30 July 2004, Heidelberg (Germany). AIP Proceedings Series (2005) astro-ph/0501576

Ramaty, R., Kozlovsky, B., Lingenfelter, R., Reeves, H., ApJ 488, 730 (1997)

Reeves, H., Reviews of Modern Physics, Volume 66, 193-216 1994

Romani, R., ApJ 399, 621 (1992)

Rathborne, J. M., Jackson, J. M., Simon, R., ApJ 641, 389 (2006)

Verner, D. & Ferland, J., ApJS, 103, 467 (1996)

White, N. E. and van Paradijs, J., ApJL 474, 25 (1996)

Woosley, S. E., Hartmann, D., Hoffman, R. D. and Haxton, W. C., Astrophysical Journal, vol. 356, p. 272-301. (1990)

Yoshida, T., Terasawa, M., Kajino, T., Sumiyoshi, K., ApJ 600, 204 (2004)

YMB is supported by NASA/Chandra and NASA/INTEGRAL General Observer Grants and a NASA Long Term Space Astrophysics Grant.

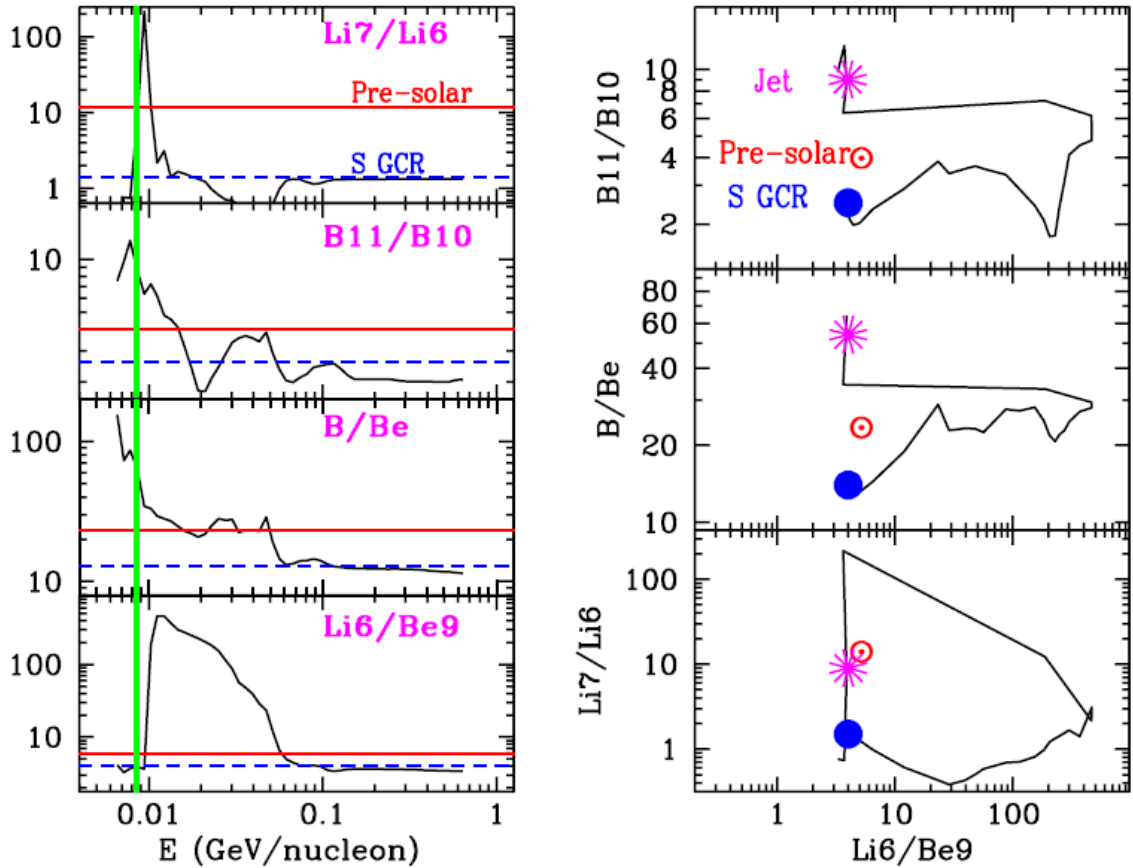

Fig. 1:

*Left:* Abundance ratios of light isotopes, produced by a monoenergetic beam of solar composition irradiating a target of solar composition, as a function of the beam energy. In all panels, horizontal solid lines indicate pre-solar (meteoritic) composition, while dashed horizontal lines indicate ratios produced by the standard galactic cosmic ray (SGCR) spectra. The green vertical line indicates the jet energy, ~8-9 MeV/n, where production of large $^{11}$B/$^{10}$B ratio is possible, *without overproduction of any other ratio*.

*Right:* Same as on the left, but abundance ratios are plotted as a function of $^6$Li/$^9$Be. The pre-solar (meteoritic), standard GCR, and Jet (8-9 MeV/n) ratios are indicated by the dot (o), filled circle and asterisk, respectively, in all panels. Note that both $^{10}$B and $^{11}$B are produced by the jet-ISM interaction and that the B/Be ratio is higher than its pre-solar value in roughly the same proportion as the $^{11}$B/$^{10}$B ratio exceeds its pre-solar value (2.34 and 2.5, respectively), so that after dilution with material submitted to only normal SGCR irradiation, we recover both the pre-solar B/Be and $^{11}$B/$^{10}$B ratios.